\documentclass[
    ,final            
  ]
  {aipproc}

\layoutstyle{6x9}
\usepackage{amssymb}
\def\apj{ApJ}%
\def\aap{A\&A}%
\def\nat{Nature}%
\def\pasp{PASP}%

\begin{document}

\title{Rotating massive stars @ very low Z: \\
high C \& N production}


\author{Raphael HIRSCHI}{
  address={Dept. of physics and Astronomy,
  University of Basel,
  Klingelbergstr. 82,
  4056 Basel,
  Switzerland}
}

\begin{abstract}
Two series of models and their yields are presented in this paper.
The  first series consists of 20 $M_\odot$ models with varying initial
metallicity (solar down to $Z=10^{-8}$) and rotation
($\upsilon_{\rm ini}=0-600$\,km\,s$^{-1}$). The second one consists of
models with an initial metallicity of $Z=10^{-8}$, masses between 20 and
85 $M_\odot$ and average rotation velocities at these metallicities 
($\upsilon_{\rm ini}=600-800$\,km\,s$^{-1}$).
The most interesting models are the models 
with $Z=10^{-8}$ ([Fe/H]$\sim-6.6$).
In the course of helium burning, carbon and oxygen are mixed into the
hydrogen burning shell. This boosts the importance of the shell and
causes a reduction of the size of the CO core. Later in the evolution,
the hydrogen shell deepens and produces large amount of primary
nitrogen. For the most massive models ($M\gtrsim 60$\,$M_\odot$),
significant mass
loss occurs during the red supergiant stage. This mass loss is due to
the surface enrichment in CNO elements via rotational and convective
mixing. 

The yields of the fast rotating 20 $M_\odot$ models can 
best reproduce (within our study) the observed abundances at the surface of extremely 
metal poor (EMP) stars.
The wind of the massive models can reproduce the CNO abundances of the
carbon--rich UMPs, in particular for the most metal poor star known to
date, HE1327-2326.
\end{abstract}

\maketitle


\section{Introduction}
Precise measurements
of abundances of extremely metal poor (EMP) stars have recently 
been obtained by \citet{FS5,FS6,IER04}, \ldots . 
These provide new constraints for the
stellar evolution models \citep[see][]{CMB05,Fr04,Pr04}.
The most striking constraint is the need for primary $^{14}$N production in very low
metallicity massive stars. Other possible constraints are an upturn of the 
C/O ratio with a [C/Fe]
about constant or slightly decreasing (with increasing metallicity) at very low metallicities, 
which requires an increase (with increasing metallicity) of oxygen yields
below [Fe/H]$\sim$ -3. 
About one quarter of EMP stars are carbon rich (C-rich EMP, CEMP stars).
\citet{RANB05} propose a classification for these stars. They find two
categories: about three quarter are main s-process enriched (Ba-rich) 
stars and one quarter are enriched with a weak component of
s-process (Ba-normal). The two most metal poor stars known to date, 
HE1327-2326 \citep{Fr05,Ao05} and HE 0107-5240 \citep{Ch04} are both
CEMP stars. These stars are believed to have been enriched by only one
to several stars and we can therefore compare our yields to
their observed abundances without the filter of a galactic chemical
evolution model (GCE).
In an attempt
to explain the origin of the abundances observed as well as the
metallicity trends, I computed
pre-supernova
evolution models of rotating single stars with metallicities ranging 
from solar metallicity down to $Z=10^{-8}$ following the work of
\citet{MEM05}. 

\section{Description of the stellar models}
The computer model used to calculate the stellar models
 is described in detail in \citet{psn04a}.
At low metallicities the mixture of the heavy elements 
we adopted is the one
used to compute the opacity tables for Weiss 95's alpha--enriched 
composition \citep{IR96}. 
The mass loss rates are described and discussed in \citet{MEM05}. 
Very little was known about the mass
loss of very low metallicity stars with a strong enrichment in CNO
elements until recently. \citet{VdK05} study the case of WR stars but a
crucial case, which has not been studied in detail yet, is the case of red
supergiant stars (RSG). As we shall see later, due to rotational and
convective mixing, the surface of the star is strongly enriched in CNO
elements during the RSG stage. Awaiting for future studies, 
it is implicitly assumed 
in this work (as in \citet{MEM05}) 
that CNO elements have a significant contribution to
opacities and mass loss rates. This assumption
is supported by the possible formation of
molecular lines in the RSG stage.
Therefore the mass loss rates depend 
on metallicity as $\dot{M} \sim (Z/Z_{\odot})^{0.5}$, where
$Z$ is the mass fraction of heavy elements at the surface
of the star.
The evolution of the models was in general followed until core Si--burning and 
the stellar yields are calculated as in \citet{ywr05}.
The main characteristics of the models are presented 
in Table \ref{table1}. More details about the models are presented in \citet{H06}.

The value of 300 km\,s$^{-1}$ used for the initial rotation velocity 
at solar metallicity
corresponds to an average velocity of about 220\,km\,s$^{-1}$ on the Main
Sequence (MS) which is
very close to the average observed value \citep[see for instance][]{FU82}. 
It is unfortunately not possible to measure the rotational velocity of 
very low
metallicity massive stars since they all died a long time ago.
Nevertheless, there is indirect evidence that stars with a lower metallicity have a
higher rotation velocity. This can be due to the
difficulty of evacuating angular momentum during the star formation, which is even more
important at lower metallicities 
\citep[see][]{ABN02}.
Furthermore, a very low metallicity star containing the same angular momentum
as a solar metallicity star has a higher surface rotation velocity due to
its smaller radius (one quarter of $Z_\odot$ radius for 20 $M_\odot$
stars).
In order to compare the
models at different metallicities and with different initial masses, 
the ratio $\upsilon_{\rm ini}/ \upsilon_{\rm crit}$ is used (see Table
\ref{table1}).
$\upsilon_{\rm crit}$ is the critical velocity at which matter becomes 
gravitationally unbound.
$\upsilon_{\rm ini}/ \upsilon_{\rm crit}$ 
 increases only as $r^{-1/2}$ for
models with the same angular momentum ($J$) but lower metallicity, 
whereas the surface rotational velocity increases as $r^{-1}$ 
($J\sim \upsilon r$).
The angular momentum can be compared as well but one has to bear in mind
that it varies significantly
for models of different initial masses.
Finally,
$\upsilon_{\rm ini}/ \upsilon_{\rm crit}$ is a good indicator for the
impact of rotation on mass loss.

In the first series of models, the aim is to scan the parameter space of
rotation and metallicity with 20 $M_\odot$ models since a 20 $M_\odot$
star is not far from the average massive star concerning stellar yields. 
For this series,
two initial rotational velocities were 
used at very low metallicities. 
The first one is the same as at solar metallicity, 
300\,km\,s$^{-1}$.
The second $\upsilon_{\rm ini}$ is 
500 at Z=10$^{-5}$ ([Fe/H]$\sim $-3.6) and 600\,km\,s$^{-1}$
at Z=10$^{-8}$ ([Fe/H]$\sim$-6.6). These second values have
ratios of the initial velocity to the break--up velocity, 
$\upsilon_{\rm ini}/\upsilon_{\rm crit}$, around 0.55, which is only slightly
larger than the solar metallicity value (0.44). 
The 20 $M_\odot$ model at Z=10$^{-8}$
 and with 600\,km\,s$^{-1}$ has a total initial angular momentum $J_{\rm
tot}=3.3\,10^{52}$\,erg\,s which is the same as the solar metallicity 20 $M_\odot$ model with 300\,km\,s$^{-1}$ ($J_{\rm
tot}=3.6\,10^{52}$\,erg\,s).
So a velocity of 
600\,km\,s$^{-1}$, which at first sight seems extremely fast, is probably
the average velocity at Z=10$^{-8}$.
In the second series of models, I follow the exploratory work of
\citet{MEM05} and compute models at Z=10$^{-8}$ with initial masses of
40, 60 and 85 $M_\odot$ and initial rotational velocities of 700, 800 and
800\,km\,s$^{-1}$ respectively. Note that, for these models as well, the
initial total angular momentum is similar to the one contained in solar 
metallicity models with rotational velocities of 300\,km\,s$^{-1}$. 
Since this is the
case, velocities between 600 and 800\,km\,s$^{-1}$ are considered in
this work as the average rotational velocities at these very low metallicities.

\begin{table}
\begin{tabular}{r r r r r r r r r r r r r }
\hline \hline 
$M_{\rm{ini}}$ & $Z_{\rm{ini}}$ & $\upsilon_{\rm{ini}}$ & $J_{\rm{tot}}^{\rm{ini}}$ & $\frac{\upsilon_{\rm ini}}{\upsilon_{\rm crit}}$
& $\tau_{\rm{life}}$ 
& $M_{\rm{final}}$ & $M_{\alpha}$ & $M_{\rm{CO}}$ & $M_{\rm{rem}}$  
& $^{12}$C & $^{14}$N & $^{16}$O  \\ 
\hline
20 & 2e-02 & 300 & 0.36 & 0.44 & 11.0 &  8.763 & 8.66 & 6.59 & 2.57 & 0.433 & 4.33e-2 & 2.57 \\
20 & 1e-03 & 000 &  --  & 0.00 & 10.0 & 19.557 & 6.58 & 4.39 & 2.01 & 0.373 & 3.31e-3 & 1.46 \\
20 & 1e-03 & 300 & 0.34 & 0.39 & 11.5 & 17.190 & 8.32 & 6.24 & 2.48 & 0.676 & 3.10e-3 & 2.70 \\
20 & 1e-05 & 000 &  --  & 0.00 & 9.80 & 19.980 & 6.24 & 4.28 & 1.98 & 0.370 & 4.27e-5 & 1.50 \\
20 & 1e-05 & 300 & 0.27 & 0.34 & 11.1 & 19.930 & 7.90 & 5.68 & 2.34 & 0.481 & 1.51e-4 & 2.37 \\
20 & 1e-05 & 500 & 0.42 & 0.57 & 11.6 & 19.575 & 7.85 & 5.91 & 2.39 & 0.648 & 5.31e-4 & 2.59 \\
20 & 1e-08 & 000 &  --  & 0.00 & 8.96 & 19.999 & 4.43 & 4.05 & 1.92 & 0.262 & 8.52e-3 & 1.20 \\
20 & 1e-08 & 300 & 0.18 & 0.28 & 9.98 & 19.999 & 6.17 & 5.18 & 2.21 & 0.381 & 1.20e-4 & 1.96 \\
20 & 1e-08 & 600 & 0.33 & 0.55 & 10.6 & 19.952 & 4.83 & 4.36 & 2.00 & 0.823 & 5.90e-2 & 1.35 \\
40 & 1e-08 & 700 & 1.15 & 0.55 & 5.77 & 35.795 & 13.5 & 12.8 & 4.04 & 1.79  & 1.87e-1 & 5.94 \\
60 & 1e-08 & 800 & 2.41 & 0.57 & 4.55 & 48.975 & 25.6 & 24.0 & 7.38 & 3.58  & 4.14e-2 & 12.8 \\
85 & 1e-08 & 800 & 4.15 & 0.53 & 3.86 & 19.868 & 19.9 & 18.8 & 5.79 & 7.89  & 1.75e+0 & 12.3 \\
\hline
\end{tabular}
\caption{Initial parameters of the models (columns 1--5): 
mass, metallicity, rotation velocity [km\,s$^{-1}$], 
total angular momentum [10$^{53}$\,erg\,s] and $\upsilon_{\rm ini}/ \upsilon_{\rm crit}$.
Total lifetime [Myr] and
various masses [$M_\odot$] (7--10): final mass, masses of the helium
and carbon--oxygen cores and the remnant mass.
Total stellar yields (wind + SN) [$M_\odot$] for carbon (11), nitrogen 
(12) and oxygen (13).}
\label{table1}
\end{table}

\section{Evolution of the 20 $M_\odot$ models}
\begin{figure}
 
\includegraphics[height=.25\textheight]{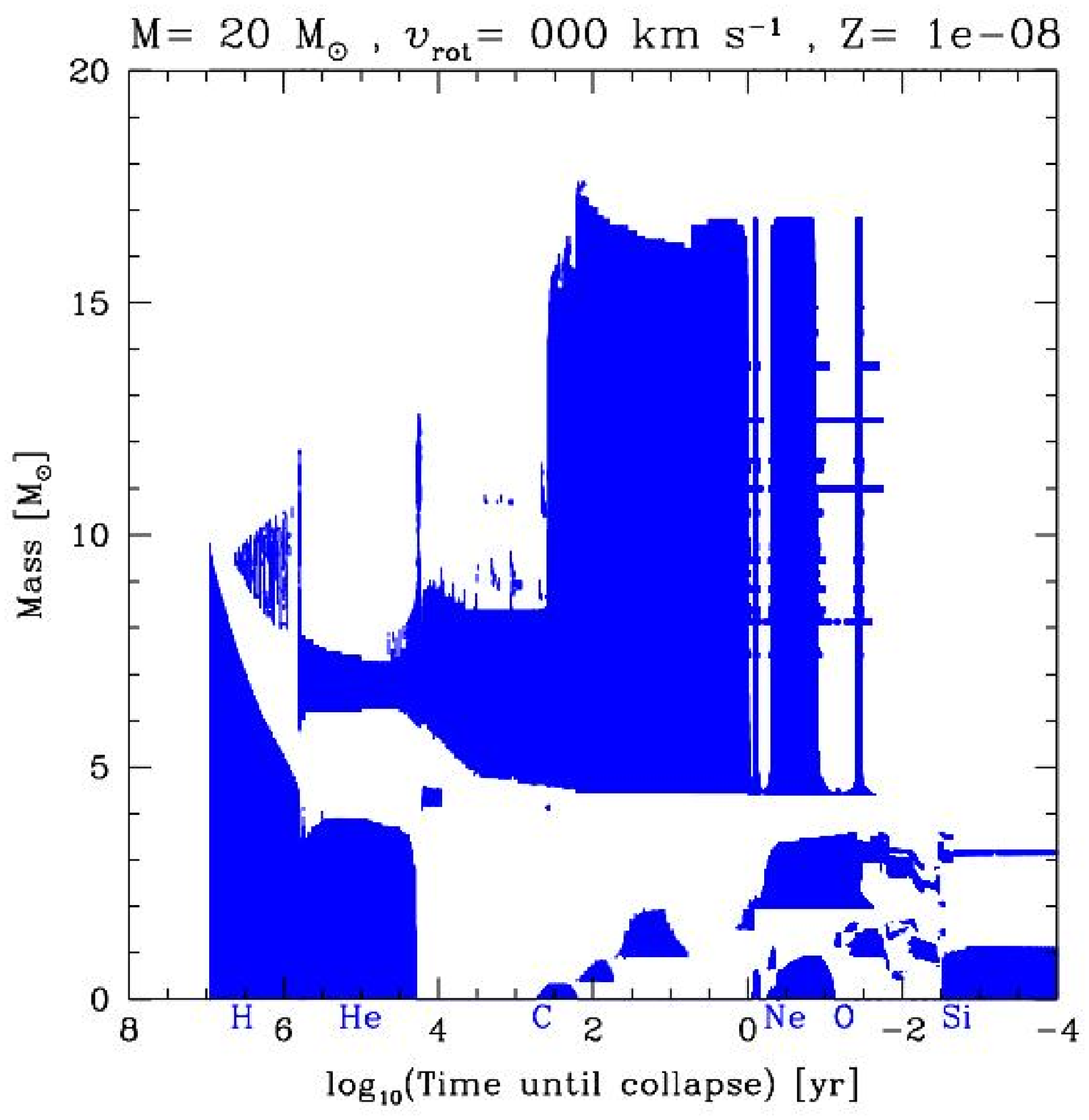}\includegraphics[height=.25\textheight]{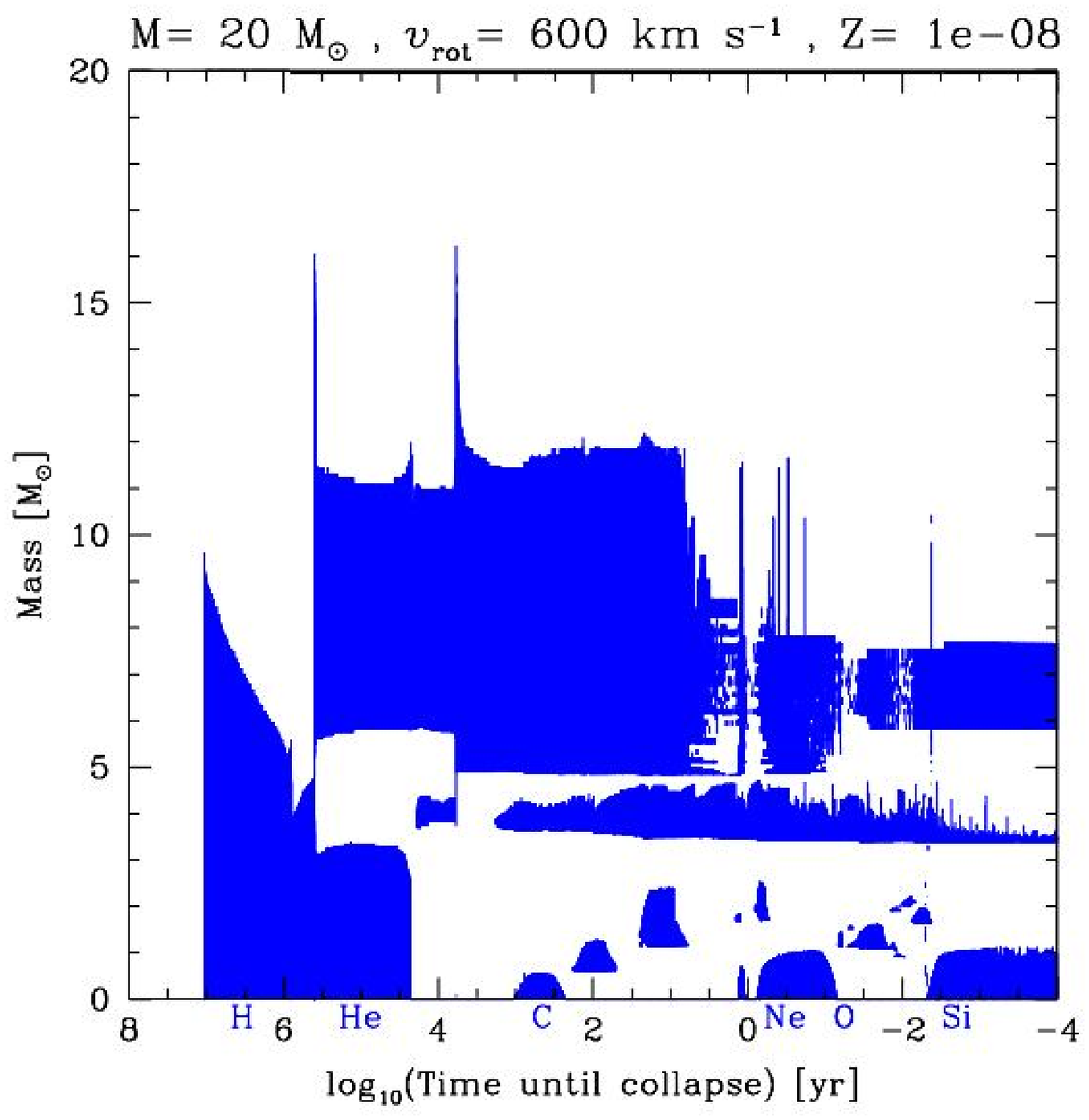}
  \caption{Kippenhahn diagrams of 20$\,M_\odot$ models 
at $Z=10^{-8}$ with $\upsilon_{\rm ini}=$
 0\,km\,s$^{-1}$ ({\it left}) and
600\,km\,s$^{-1}$ ({\it right}).}
\label{kip20}
\end{figure}
Mass loss becomes gradually unimportant as the metallicity decreases in
the 20 $M_\odot$ models. At solar metallicity, the rotating 20 $M_\odot$
model loses more than half of its mass and at 
$Z=10^{-8}$ less than 0.3\% (see Table \ref{table1}). This means that 
at very low metallicities,
the dominant effect of rotation is mixing for the mass range around 20
$M_\odot$. 
The impact of rotational mixing is best
pictured in the Kippenhahn diagram (see Fig. \ref{kip20}). 
During hydrogen burning and the start of helium burning, mixing
increases the core sizes. Mixing of helium above the core suppresses
the intermediate convective zones linked to shell H--burning. So far the
impact of mixing at $Z=10^{-8}$ is the same as at higher metallicities. 
However, after some
time in He--burning, the mixing of primary carbon and oxygen into the
H--burning shell is important enough to boost significantly the strength
of the shell. As a result, the size of the helium
burning core becomes and remains smaller than in the non--rotating
model. The yield of $^{16}$O being closely
correlated with the size of the CO core, it is therefore  
reduced due to the strong mixing. At the same time carbon yields are increased.
This produces an upturn of C/O at very low metallicities.

\subsection{Stellar yields of CNO elements}

\begin{figure}
 
\includegraphics[height=.2\textheight]{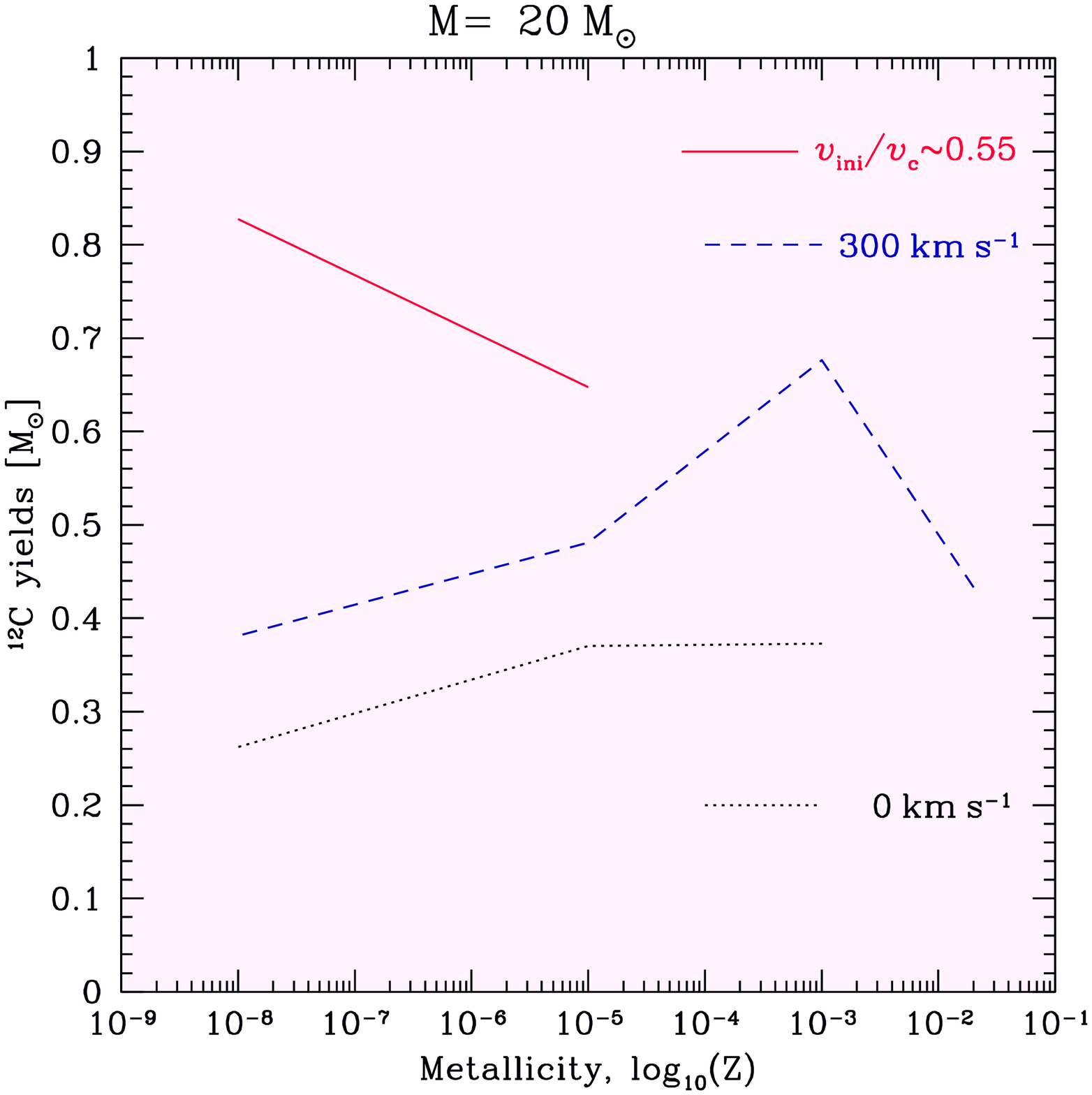}\includegraphics[height=.2\textheight]{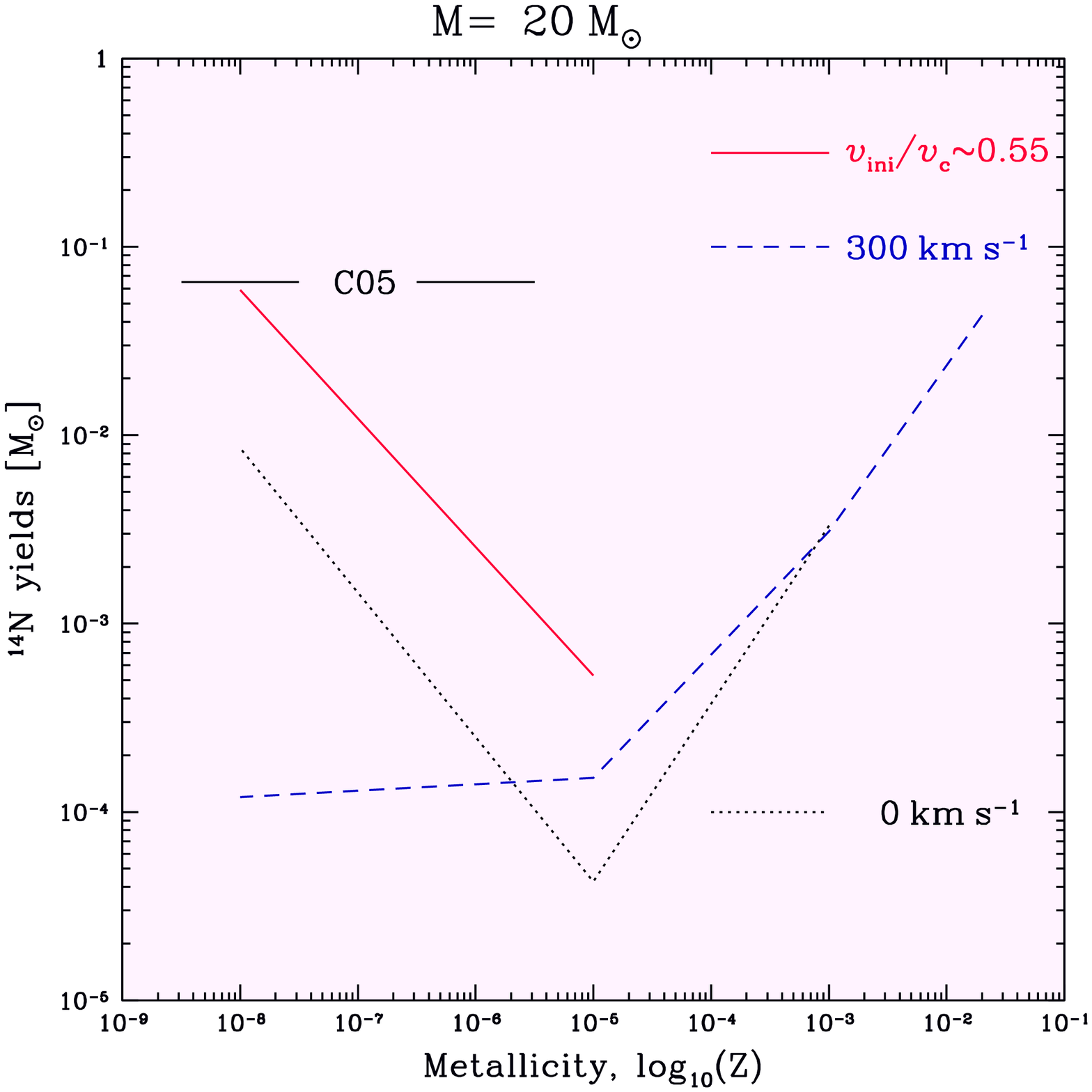}\includegraphics[height=.2\textheight]{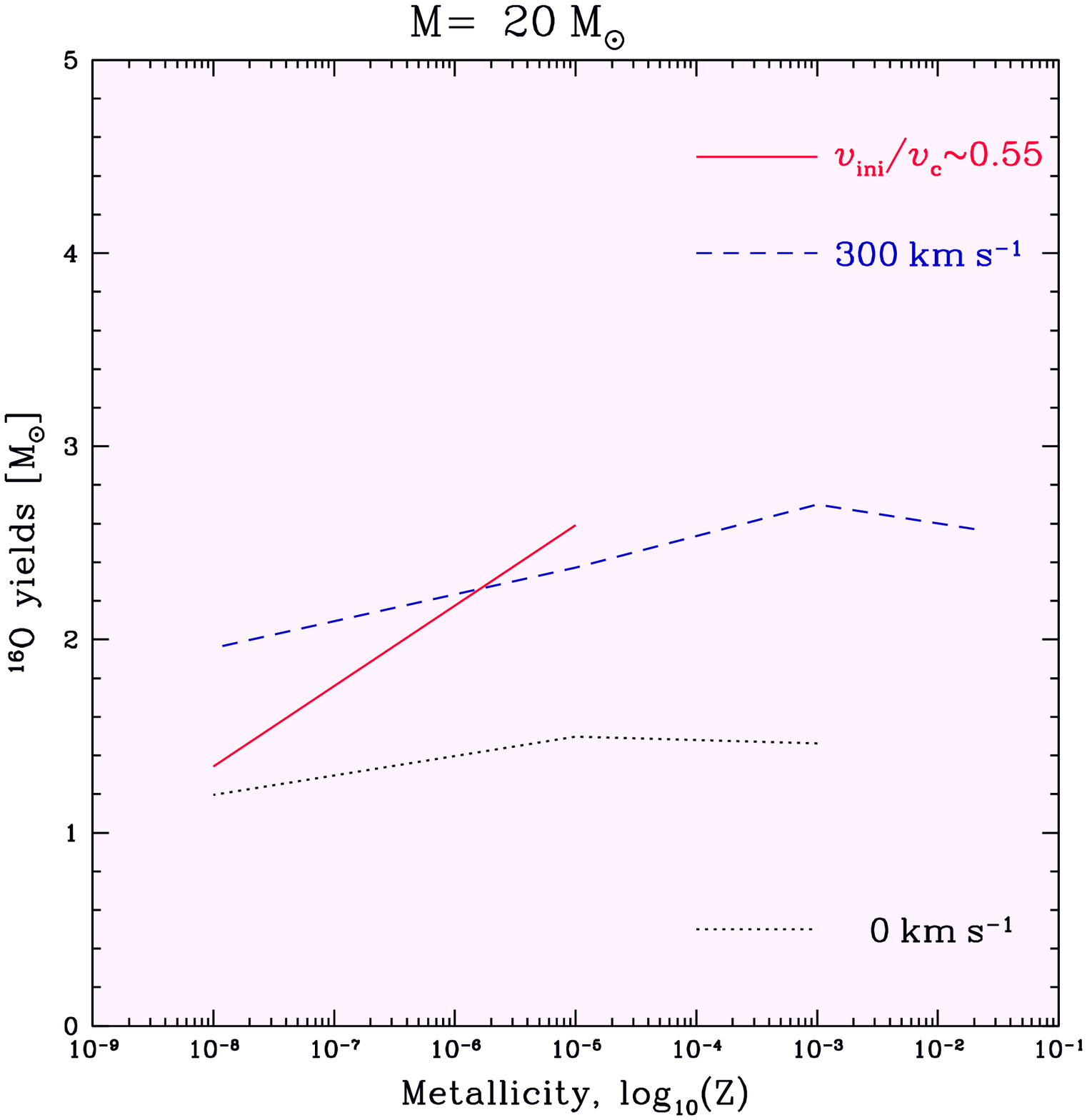}
  \caption{Stellar yields of $^{12}$C ({\it left}), $^{14}$N ({\it center}) 
and $^{16}$O ({\it right})
as a function of the initial metallicity of the models.
The solid red, dashed blue and dotted black lines represent respectively
the models with $\upsilon_{\rm ini}/\upsilon_c \sim 0.55$ 
($\upsilon_{\rm ini}$=500\,km\,s$^{-1}$ at $Z=10^{-5}$ 
and $\upsilon_{\rm ini}$=600\,km\,s$^{-1}$ at $Z=10^{-8}$),
with $\upsilon_{\rm ini}$=300\,km\,s$^{-1}$
and without rotation.
For nitrogen, the horizontal mark with C05 in the middle
corresponds to the value deduced from the chemical evolution models of
\citet{CMB05}.}
\label{ycno}
\end{figure}
The yields of $^{12}$C, $^{14}$N and $^{16}$O are presented in Fig. 
\ref{ycno} and their numerical values are given in Table
\ref{table1}
\citep[see][ for more details]{H06}.
The most stringent observational constraint at very low Z is a very
high primary $^{14}$N production \citep{CMB05,Pr04},
 of the order of 0.06 $M_\odot$ per star. 
In Fig. \ref{ycno} ({\it center}), we can see that only the model at $Z=10^{-8}$ and with
$\upsilon_{\rm ini}$=600\,km\,s$^{-1}$ can reach such high values.
The bulk of $^{14}$N is produced in the convective 
zone created by shell hydrogen burning (see Fig. 1 
{\it right}). If this 
convective zone deepens enough
to engulf carbon (and oxygen) rich layers, then significant amounts of 
primary
$^{14}$N can be produced ($\sim$0.01$\,M_\odot$). 
This occurs in both the non--rotating model 
and the fast rotating model but for different reasons.
In the non--rotating model, it occurs due to
structure rearrangements similar to the third dredge--up at the end of 
carbon burning. In the model with $\upsilon_{\rm ini}=$600\,km\,s$^{-1}$ 
it occurs during shell helium burning 
because of the strong mixing of carbon and oxygen into the 
hydrogen shell burning zone.

Models with higher initial masses at $Z=10^{-8}$ also produce large
quantities of primary nitrogen. More computations are necessary to see
over which metallicity range the large primary production takes place and
to see whether the scatter in yields of the models with different masses
and metallicities is compatible with the observed scatter.

\section{Evolution of the models at $Z=10^{-8}$}
\begin{figure}
 
\includegraphics[height=.25\textheight]{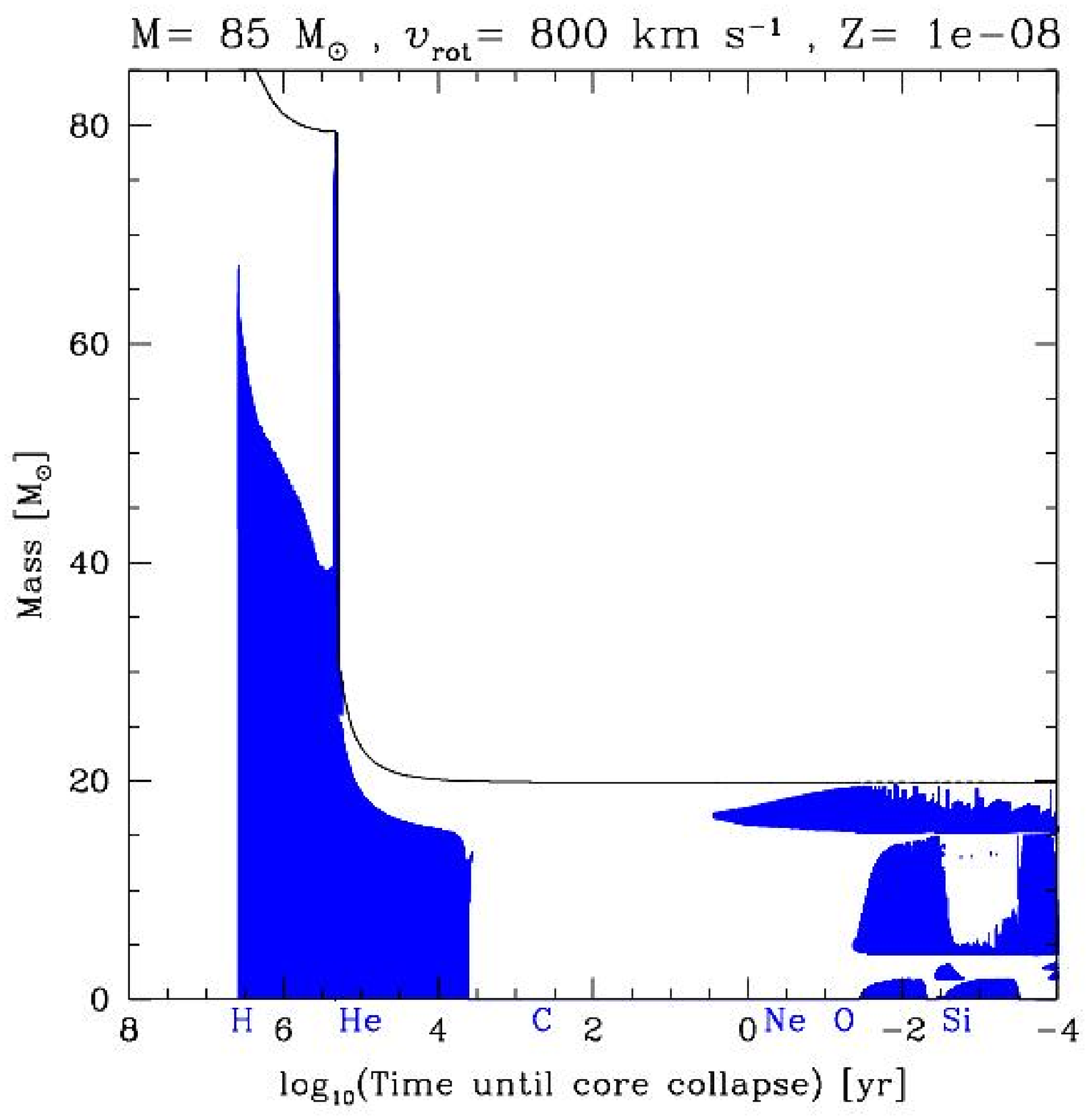}\includegraphics[height=.25\textheight]{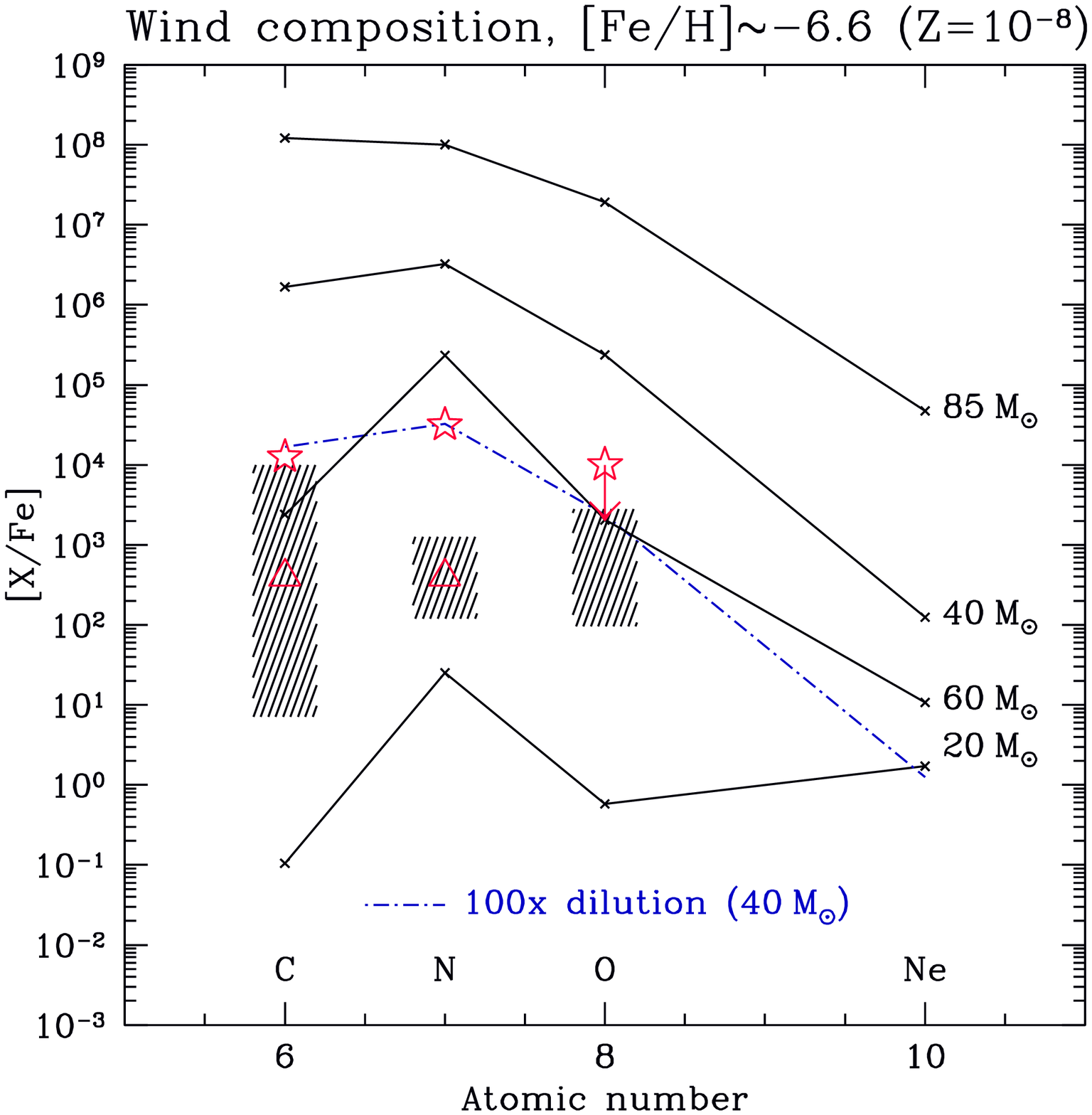}
  \caption{{\it Left}: Kippenhahn diagrams of the 85$\,M_\odot$ model 
at $Z=10^{-8}$ with $\upsilon_{\rm ini}=$
 800\,km\,s$^{-1}$. 
 {\it Right}: 
 The solid lines represent the chemical composition of
 the wind material of the of the different models at
 $Z=10^{-8}$.
The hatched areas correspond to the range of values
measured at the surface of giant CEMP stars: HE 0107-5240, [Fe/H]$\simeq$~-5.3 
\citep{Ch04};
CS 22949-037, [Fe/H]$\simeq$~-4.0 \citep{NRB01,FS02};
CS 29498-043, [Fe/H]$\simeq$~-3.5 \citep{Ao04}. 
The empty triangles \citep{PC05}([Fe/H]$\simeq -4.0$) 
and stars \citep{Fr05} ([Fe/H]$\simeq -5.4$, only an
upper limit is given for [O/Fe]) correspond to
non-evolved CEMP stars.
}
\label{m85}
\end{figure}
\begin{table}
\begin{tabular}{r r r r r r r r r r }
\hline \hline 
$M_{\rm{ini}}$ & $Z_{\rm{ini}}$ & $\upsilon_{\rm{ini}}$ 
& $^{12}$C & $^{14}$N & $^{16}$O  \\ 
\hline
20 & 1e-08 & 600 & 3.44e-12 & 3.19e-10 & 6.69e-11 \\
40 & 1e-08 & 700 & 5.34e-03 & 3.63e-03 & 2.42e-03 \\
60 & 1e-08 & 800 & 1.80e-05 & 6.87e-04 & 5.49e-05 \\
85 & 1e-08 & 800 & 6.34e+00 & 1.75e+00 & 3.02e+00 \\
\hline
\end{tabular}
\caption{Initial mass (1), metallicity (2) and rotation velocity 
[km\,s$^{-1}$] (3) and 
stellar wind ejected masses [$M_\odot$] for carbon (4), nitrogen 
(5) and oxygen (6).}
\label{wind}
\end{table}
Contrarily to what was initially expected from very
low metallicity stars, mass loss can occur in massive
stars \citep{MEM05}. The mass loss occurs in two
phases. The first phase is when the star reaches
break--up velocities towards the end of the main sequence. Due to
this effect stars, even metal free ones, are expected
to lose about 10\% of their initial masses for an
average initial rotation. The
second phase in which large mass loss can occur is
during the RSG stage. Indeed, stars more massive than
about 60 $M_\odot$ at $Z=10^{-8}$ become RSG and
dredge--up CNO elements to the surface. This brings
the total metallicity of the surface to values within
an order of magnitude of solar and triggers large mass
loss. The final masses of the models are given in
Table \ref{table1}. The case of the 85 $M_\odot$ model
is extremely interesting (see Fig. \ref{m85} {\it
left}) since it loses more than three quarter of its
initial mass. It even becomes a WO star.
\subsection{Wind composition and CRUMPS stars}
In Fig. \ref{m85} ({\it right}), we compare the
chemical composition of the wind material with
abundances observed in non-evolved carbon rich
extremely and ultra \citep{Fr05} metal poor stars. The
ejected masses of the wind material are also given in
Table \ref{wind}. It is very interesting to see that
the wind material can reproduce the observed abundance
in two ways. Either, the wind material is richer than
necessary and dilution (by a factor 100 for example
for the 40 $M_\odot$ models and HE1327-2326) with the ISM is needed or the wind 
has the right enrichment (for example the 60 $M_\odot$
and HE1327-2326) and the low mass star could form from
pure wind material. The advantage of the pure wind
material is that it has a ratio $^{12}$C/$^{13}$C around
5 \citep{MEM05} and it can explain Li depletion.
With or without dilution, the wind material
has the advantage that it brings the initial
metallicity of the low mass star above the critical
value for its formation \citep{Br05}.


\begin{theacknowledgments}
I would like to thank warmly the organisation and Prof.
Nomoto and his group for the financial support and the kind hospitality.
\end{theacknowledgments}

\bibliographystyle{aipproc}   


\end{document}